\documentclass{jps-cp}
\usepackage{txfonts} 

\title{ALICE ITS Upgrade for LHC Run 3: Commissioning in the Laboratory}

\author{\textsc{Domenico Colella}$^{1}$ for the ALICE Collaboration}

\inst{$^{1}$ Wigner Research Center for Physics, Konkoly-Thege Miklós út 29-33, 1121 Budapest, Hungary }

\email{domenico.colella@cern.ch}

\recdate{November 3, 2020}

\abst{ALICE is the CERN LHC experiment optimised for the study of the strongly interacting matter produced in heavy-ion collisions and devoted to the characterisation of the quark-gluon plasma. To achieve the physics program for LHC Run 3, a major upgrade of the experimental apparatus is ongoing. A key element of the upgrade is the substitution of the Inner Tracking System (ITS) with a completely new silicon-based detector (ITS 2) whose features will allow the reconstruction of rare physics channels not accessible with the previous layout. The enabling technology for such a performance boost is the adoption of custom-designed CMOS Monolithic Active Pixel Sensor as detecting elements. In this proceedings, an overview of the adopted technologies as well as the status of the detector commissioning will be given.}

\kword{Tracking System, MAPS}

\begin{document}
\maketitle

\section{ALICE Experiment upgrade}
ALICE (A Large Ion Collider Experiment) \cite{JINST} is a general-purpose, heavy-ion experiment at the CERN LHC. Its main goal is to study the physics properties of the quark-gluon plasma (QGP). During LHC Run 1 and Run 2, the analysis of  the data collected in heavy-ion collisions allowed the observation of hot hadronic matter at unprecedented values of temperatures, densities and volumes. These studies confirmed the basic picture, emerged from the experimental investigation at lower energies, of a QGP as an almost inviscid liquid. Moreover, precision and kinematic reach of all significant probes of the QGP were extended with respect to previous measurements. The study of the strongly-interacting state of matter in the second generation of LHC heavy-ion studies in Run 3 and Run 4 will focus on rare processes such as production of heavy-flavour particles, quarkonium states, real and virtual photons and heavy nuclear states \cite{ALICEupLoI}. The earlier methods of triggering will be limited for many of these measurements, particularly at low-$p_{\rm T}$. Therefore, the ALICE collaboration planned to upgrade the Run 1/Run 2 detector by enhancing its low-momentum vertexing and tracking capability, allowing data taking at substantially higher rates and preserving the already remarkable particle identification capabilities.
The upgraded experimental apparatus is designed to readout all \mbox{Pb--Pb} interactions, accumulating events corresponding to an integrated luminosity of more than 10 $nb^{-1}$. This minimum-bias data sample will provide an increase in statistics by about a factor 100 with respect to data collected during Run 1/Run 2 data taking. In summary, the ALICE upgrade consists of the following sub-system upgrades:
\begin{itemize}
\item Reduction of the beam-pipe radius from \mbox{29.8 mm} to \mbox{19.2 mm}. 
\item Installation of two new high-resolution, high-granularity, low material budget silicon trackers:
  \begin{itemize}
  \item Inner Tracking System (ITS 2) \cite{ITSUPTDR} in the central pseudo-rapidity ($-1.22 < \eta < 1.22$).
  \item Muon Forward Tracker \cite{MFTTDR} covering forward pseudo-rapidity ($-3.60 < \eta < -2.45$).
  \end{itemize}
\item Replacement of the endcap wire chambers of the Time Projection Chamber by GEM detectors and installation of new readout electronics allowing continuous readout \cite{TPCUPTDR}.
\item Upgrades of the forward trigger detectors and the Zero Degree Calorimeter \cite{RDOUPTDR}.
\item Upgrades of the readout electronics of the Transition Radiation Detector, Time-Of-Flight, Photon Spectrometer and Muon Spectrometer for high rate operation \cite{RDOUPTDR}.
\item Upgrades of online and offline systems (O$^{2}$ project) \cite{O2TDR} in order to cope with the expected data volume.
\end{itemize}

\section{Inner Tracking System during Run 3}
\begin{figure}[t]
\centering
\includegraphics[width=37pc,clip]{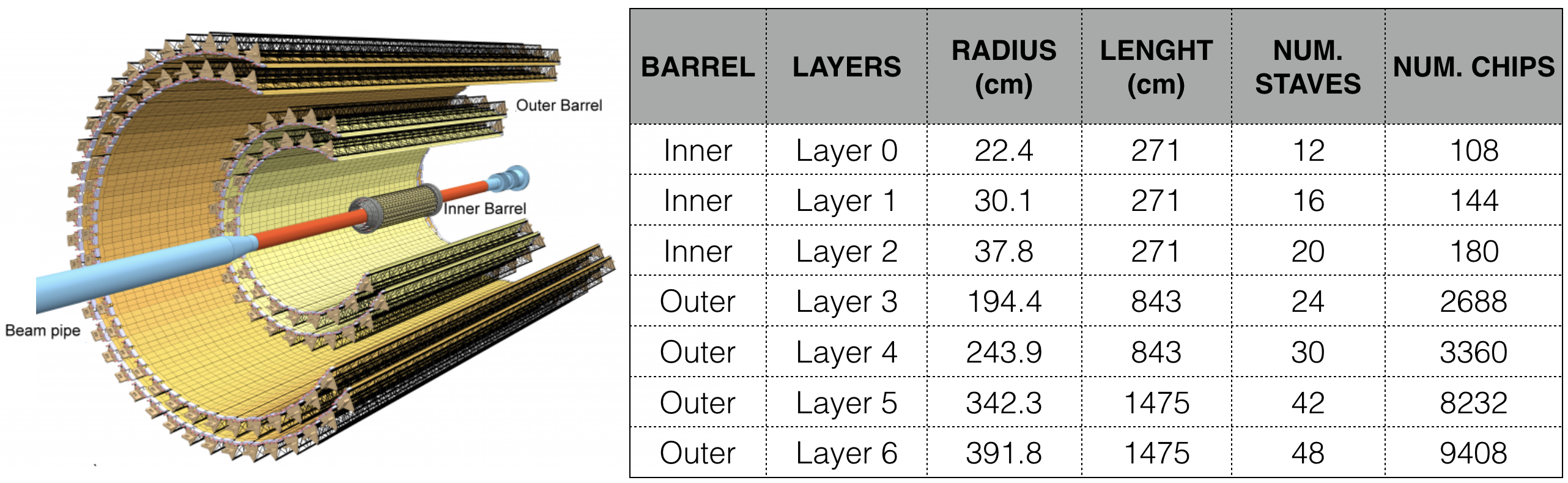}
\caption{(Left) Schematic layout of the upgraded ITS. (Right) Summary table containing the detector layers geometric characteristics.}
\label{fig-1}
\end{figure}

The main goals of the ITS upgrade (ITS 2) are to achieve an improved reconstruction of the primary vertex as well as decay vertices originating from heavy-flavour hadrons, and an improved performance for the detection of low-$p_{\rm T}$ particles. 
The design objectives are to improve the impact-parameter resolution by a factor of 3 and 5 in the r$\phi$- and z-coordinate, respectively, at a $p_{\rm T}$ of \mbox{500 MeV/c} \cite{ITSUPTDR}. The tracking efficiency and the $p_{\rm T}$ resolution at low $p_{\rm T}$ will also improve. Additio\-nal\-ly, the readout rate will be increased to \mbox{50 kHz} in \mbox{Pb--Pb} and \mbox{400 kHz} in \mbox{pp} collisions.
In order to achieve this performances, the following measures were taken:
\begin{itemize}
\item Granularity increased by an additional seventh layer and by equipping all layers with pixel sensors having cell size of  \mbox{29.24 $\mu$m $\times$ 26.88 $\mu$m}. 
\item Innermost detector layer moved closer to the interaction point, from \mbox{39 mm} to \mbox{22.4 mm}.
\item Material budget reduced down to 0.3\% X$_{0}$ per layer for the innermost layers and to 1.0 \% X$_{0}$ for the outer layers.
\end{itemize}
A schematic representation of the ITS 2 is shown in Fig. \ref{fig-1}. Based on geometrical position and few mechanical characteristics, detector can be pictured as assembled in two parts: the innermost three layers form the Inner Barrel (IB); the middle two and the outermost two layers form the Outer Barrel (OB). 
Layer radial positions (listed in table in Fig. \ref{fig-1}) were optimised to achieve the best combined performance in terms of pointing resolution, $p_{\rm T}$ resolution and tracking efficiency in the expected high track-density environment of a \mbox{Pb--Pb} collision. 
The detector pseudo-rapidity coverage is $|\eta| < 1.22$ for 90\% of the most luminous beam interaction region, extending over a total surface of 10.3 m$^{2}$ and containing about \mbox{12.5 $\times$ 10$^{9}$} pixels with binary readout. It is operated at room temperature, which is maintained between 20 $^{\circ}{C}$ and 30 $^{\circ}{C}$ using water cooling. The radiation load at the innermost layer is expected to be 270 krad of Total Ionising Dose (TID) and \mbox{1.7 $\times$ 10$^{12}$ 1 MeV n$_{eq}$/cm$^{2}$} of Non-Ionising Energy Loss (NIEL). In order to meet the material budget requirements, the silicon sensors are thinned down to \mbox{50 $\mu$m} and to \mbox{100 $\mu$m} in the IB and OB, respectively.

\subsection{Components and Layout}
The ITS 2 will be the first large-area silicon tracker based on the CMOS Monolithic Active Pixel Sensor (MAPS) technology operating at a collider. The ALPIDE, the pixel chip developed for the ITS 2, has been designed to fulfil the requirements of both the IB and OB. It is manufactured using the TowerJazz \mbox{180 nm} CMOS Imaging Sensor Process \cite{TJ}, has dimensions \mbox{15 $\times$ 30 mm$^{2}$} and hosts \mbox{512 $\times$ 1024} pixels. The final pixel chip ALPIDE was submitted in spring 2016. 
Each pixel cell contains a sensing diode, a front-end amplifier and shaping stage, a discriminator, as well as a multi-hit buffer. The pixels are arranged in double columns and read out by a priority encoder which sends the addresses of the pixels that recorded a hit in the chip peripheral readout circuitry \cite{ALPIDE2}. Pixels with no hits are not read out, making the readout procedure faster and reducing the power consumption, which for every pixel is about \mbox{40 nW}. In triggered mode, a short strobe of the order \mbox{100 ns} is used to store the hits in the in-pixel memories. In continuous-integration mode, the strobe is active for the full integration window (of the order \mbox{10 $\mu$s}) except for the switch over (about \mbox{100 ns}) from one to the other in-pixel hit buffer. 
There are several parameters, adjustable via on-chip 8-bit digital-to-analogue converters (DAC), that allow fine tuning of signal shaping and gain. The largest impact on the discrimination threshold is obtained by adjusting the parameter current threshold, I$_{THR}$. Detection efficiency has been measured as a function of this parameter, and found to be above the required 99\% for a large range of values (30 to \mbox{140 e$^{-}$}) \cite{ALPIDE1}. In the same range, the fake-hit rate stays at the sensitivity limit (\mbox{10$^{-10}$ hits/pixel/event}), after masking 0.015\% of the pixels. After NIEL irradiation up to the ten times nominal value, the detection efficiency drops below 99\% at about \mbox{120 e$^{-}$}, while the fake-hit rate is not influenced by NIEL irradiation. 
The TID irradiation ($\sim$350 krad) leads to an effective change of charge threshold of the front-end circuitry resulting in a higher fake-hit rate at low I$_{THR}$ and detection efficiency close to 100\% over the full range. In laboratory it was verified that a second internal DAC parameter, V$_{CASN}$, allows to compensate for the above mentioned effective change of charge threshold. As a consequence, the operational range in terms of detection efficiency and fake-hit rate is not reduced. 

The basic element of a layer is the stave, which consists of a carbon space frame, supporting the chips, to which the cold plate and the cooling ducts are attached. Pixel chips (9 for the IB and 14 for the OB) glued to a Flexible Printed Circuits (FPC) constitute the Hybrid Integrated Circuit (HIC). The HIC(s) are then glued to the space frame: 1 HIC for the IB and 8 or 14 HICs, actually subdivided in two adjacent half-staves, for the two innermost and the two outermost layers of the OB, respectively. The FPC consists of a polyimide with a low thermal expansion coefficient plus aluminium and copper as conductor for the IB and OB, respectively. The chips are electrically connected to the FPC, for powering and data stream, using wire bonding to pads distributed over the entire chip surface. A power bus distributes the power to the HICs on the OB (half-)stave.

The data readout chain is segmented in three stages. The digital periphery of the ALPIDE chip produces the digitized and zero-suppressed hit data and sends them to the Readout Units (RUs) through high-speed serial data links running at either \mbox{1.2 Gbps} for the IB sensors or \mbox{400 Mbps} for the OB sensors. Chip periphery interfaces also with a clock input, and a bi-directional control bus for configuration, control, and monitoring. The RUs, FPGA-based boards \cite{ITSRUJo}, are located outside the detector acceptance in a moderate radiation environment of less than 10 krad of TID. The RUs receive the triggers from the Central Trigger Processor (CTP) via optical link and ship the data to the Common Readout Units housed in the First Level Processor computers \cite{O2TDR}, located in control rooms, where the event reconstruction starts. The same data path is used to control the detector. Power to the stave is provided through a Power Board (PB), generating two \mbox{1.8 V} supply voltages (for the analog and digital circuits of the pixel chip) and a negative voltage output for the reverse bias. The RUs and PUs are powered by a CAEN distribution system based on EASY3000 crates.

\subsection{Components production and detector assembly}
The electrical and functional tests of the ALPIDE chips were performed at CERN (CH) for the IB \mbox{50 $\mu$m} thinned chips and at Yonsei and Pusan/Inha (KR) for the OB \mbox{100 $\mu$m} thinned chips, between September 2017 and May 2019. A total of almost 4$\times{10^{4}}$ chips were tested with a detector-grade quality yield of 63.7$\%$.

The full production of HICs and staves for the IB was carried out at CERN and completed in middle of 2019. A total of 95 staves, enough to build two copies of the three inner barrel layers, were assembled with a yield of 73$\%$. The OB HICs were produced in five sites (Bari (IT), Liverpool (UK), Pusan/Inha, Strasbourg (FR) and Wuhan(CN)), while the FPCs were tested in Trieste and Catania (IT). It took 80 weeks to assembly the needed 2500 HICs; this quantity includes the HICs needed to cover the whole detector acceptance plus spares and assumes an overall production yield of 74$\%$ (convolution of 82$\%$ for the HIC and 90$\%$ for the stave). The HIC production was completed during September 2019, achieving a yield of 84$\%$. The OB staves were assembled in five sites (Berkeley (US), Daresbury (UK), Frascati (IT), NIKHEF (NL) and Torino (IT)) by October 2019, reaching a yield close to 90$\%$ for enough objects, qualified as detector grade, to cover the full OB acceptance. Additional spare staves were also assembled. Accelerated ageing tests both for HICs and staves, verified the assembly procedure and validated the use of the appropriate components \cite{AGEING}.

Production and qualification of the full set of 192 RUs (CERN, Bergen (NO), NIKHEF) and 142 PBs (Berkeley), plus spares, were completed by mid 2019.

\begin{figure}[t]
\centering
\includegraphics[width=37pc,clip]{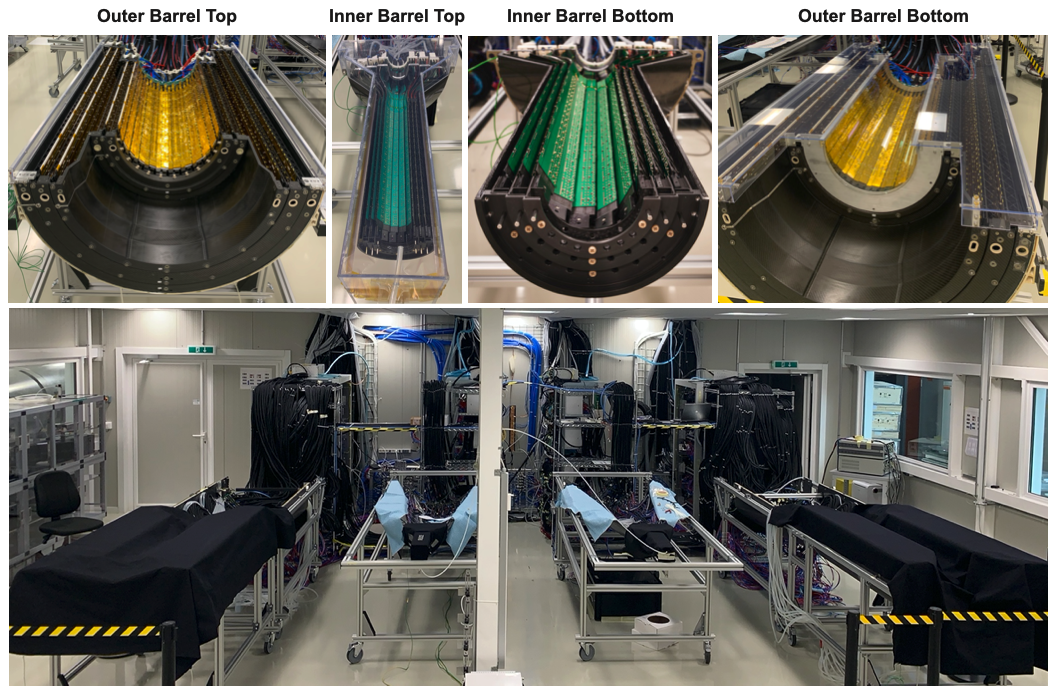}
\caption{Picture of the CERN clean room where the detector is seating while in commissioning at the surface, before installation at LHC P2 cavern. Each of the four half-barrel is connected to the corresponding crate, hosting the power and readout units.}
\label{fig-2}
\end{figure}

A large clean room (Fig. \ref{fig-2}) was built at CERN to allow the full detector assembly and the on-surface commissioning activities, before the installation in the ALICE cavern starting from January 2021. Here the same backend system that will be used in the experiment is available, including powering system, cooling system, full readout and trigger chains. Integration of the staves in the layer structure and connection to the services progressed with staves availability at CERN. Layers are assembled in half-barrels and are currently divided in 4 parts: two (top and bottom) of the IB (layers 0,1 and 2) and two (top and bottom) of the OB (layers 3,4,5 and 6). Each of them is connected to a crate hosting RUs and PBs. First completely assembled and connected half-barrel has been the IB top by July 2019; last one has been the OB bottom by January 2020.

\subsection{Commissioning in Laboratory}
Installation of the different detector components in the clean room was a long process. 
Commissioning activities become more and more specific with the availability of hardware and development of firmware/software, starting from basic functionality verifications to detailed and quantitative measurements of system performance.
First studies of the behaviour of staves assembled in layer using final powering and readout schema were performed using a lower quality half-layer 0 (Debug Layer). Continuous operation of the detector started in May 2019, supported by crews of two shifters alternating along the day every 8 hours. Coordinated by shift leader, the shift crew collects the data, assess its quality and monitors the status of the detector. Standard data taking schedule foresees hourly execution of a threshold scan, a fake-hit rate run  and a readout test. 

As anticipated, the charge threshold for pixel hit generation can be modified acting on two internal DACs (I$_{THR}$ and V$_{CASN}$) \cite{ALPIDE1}. Dedicated calibration runs are executed every time the powering schema of the chips is changed. The threshold value is always a trade-off between pixel efficiency and amount of noise. It has been measured that a good target threshold value for calibration is \mbox{100 e$^{-}$}. The threshold scan allows to verify the correctness and uniformity over the pixel chip matrix of the detector calibration. A fake-hit rate run allows to measure the amount of noise. A readout test is used to verify the quality of the data transmission between the chips and the RUs. 

Most of the activities of the shifters have been performed on the different IB pieces available with time, starting from the Debug Layer to the final full IB (top and bottom parts). OB took longer for the final assembly and required more time for the complete integration in the system, due to larger number of elements. Most of the verifications performed on the OB staves have been done by experts. 

First versions of Detector Control System (DCS) and Data Acquisition System (DAQ) as well as Quality Control (QC) and track reconstruction softwares are available and running on machines housed in a control room adjacent to the clean room. 

\subsubsection{Inner Barrel}
\begin{figure}[t]
\centering
\includegraphics[width=37pc,clip]{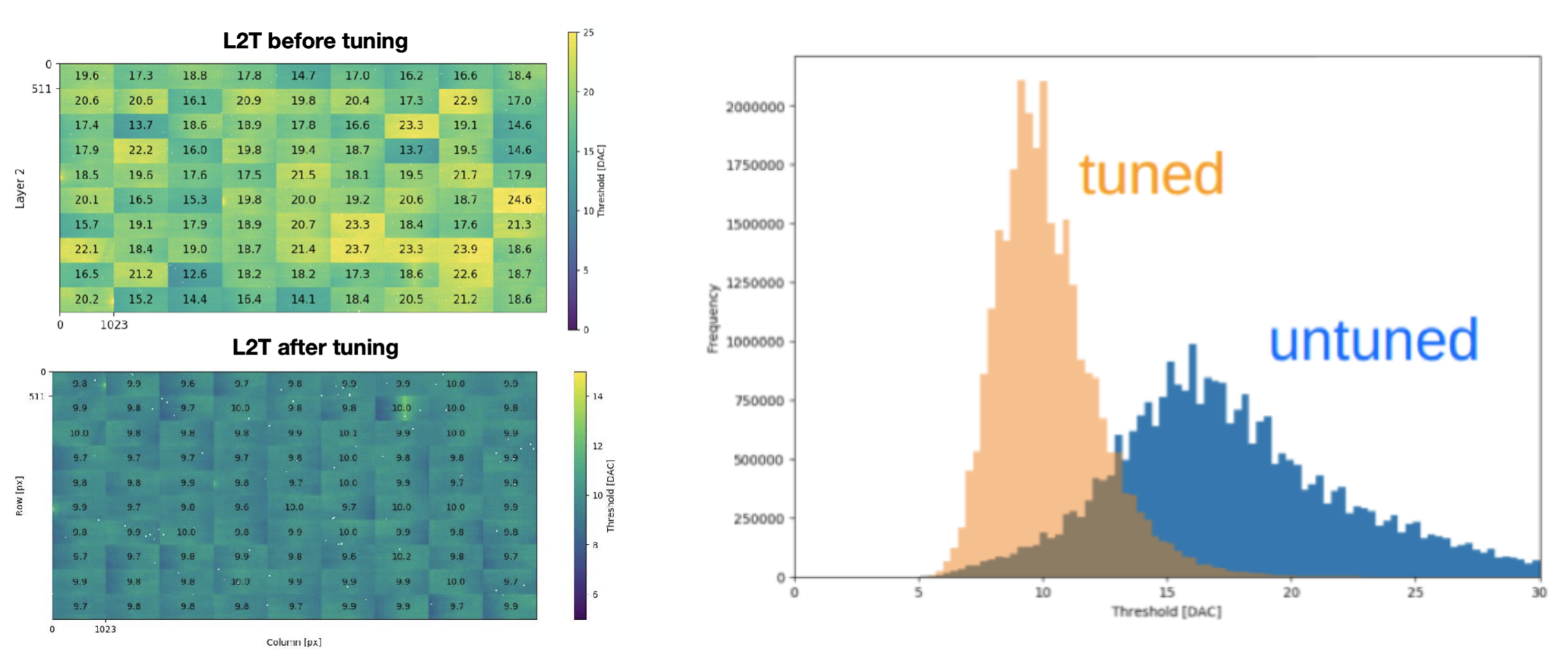}
\caption{Internal DAC tuning effectiveness: pixel threshold maps (left) and threshold value distribution (right) before and after the tuning. Threshold values are expressed in DAC units; conversion factor to e$^{-}$ is 10. Results obtained for the Layer 2 top (L2T).}
\label{fig-3}
\end{figure}

One of the first tasks was the verification of the threshold calibration procedure. Effectiveness of the DAC tuning can be determined by looking at the change of the pixel threshold distribution before and after the calibration step as in Fig. \ref{fig-3}. As anticipated, target threshold value from the calibration is \mbox{100 e$^{-}$}; values reported in the plots are expressed in DAC units, and the conversion factor to e$^{-}$ is 10. The level of homogeneity reached between the different chips in the layer can also be seen from the same figure. The large cumulated statistics of more than 5000 runs, recorded over tens of hours of running, have confirmed that the threshold value after calibration remained stable in time. This has been verified both with back-bias supply voltage equal to \mbox{0 V} and \mbox{-3 V}.  

\begin{figure}[t]
\centering
\includegraphics[width=37pc,clip]{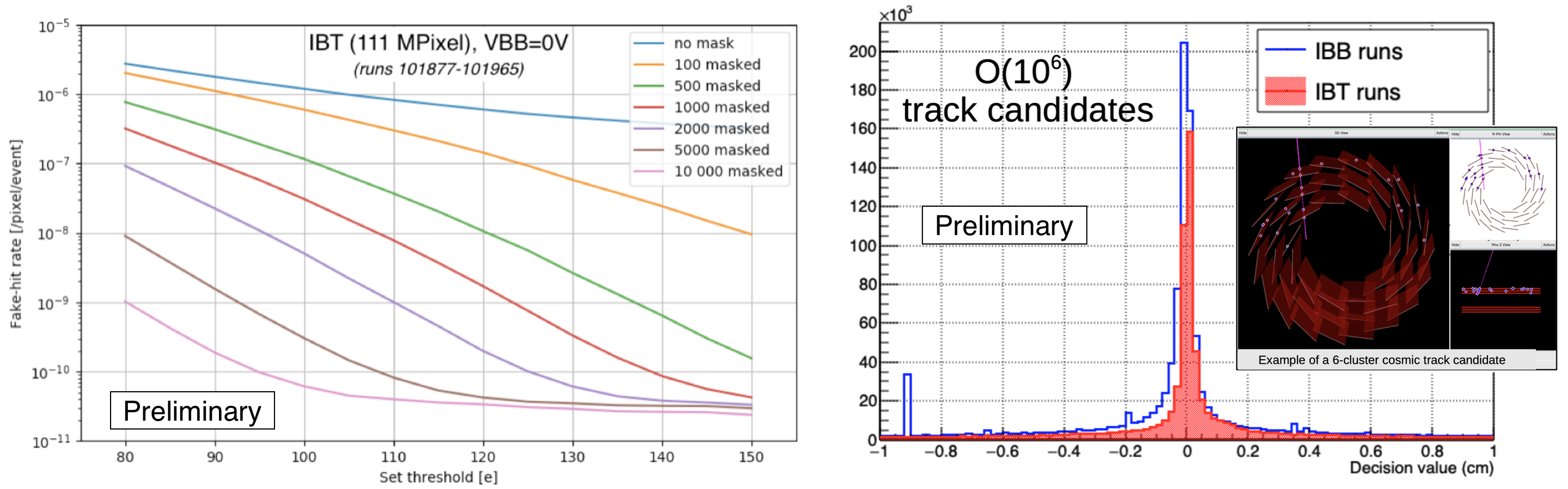}
\caption{(Left) Fake-hit rate as a function of the threshold value for different numbers of masked pixels. Results obtained for the full IB top. (Right) Cluster correlation distribution from cosmics tracks. In the event display, an example of tracks reconstructed from 6 aligned hits, obtained thanks to the overlap of the chips at the edges.}
\label{fig-4}
\end{figure}

During fake-hit rate runs, random triggers are sent to the chips and all the hits created in coincidence are collected. These hits come from the noise and also from cosmic rays crossing the detector in coincidence with the trigger. In the left plot of Fig. \ref{fig-4}, the measured fake-hit rate is plotted as a function of the threshold value for different numbers of noisier masked pixels. It can be seen that, for the target threshold value of \mbox{100 e$^{-}$}, masking a total of 10000 pixels, less than the 0.009\% of the considered acceptance, a fake-hit rate value of the order of \mbox{10$^{-10}$ hits/pixel/event} can be achieved. This is an important result, considering the really low fraction of masked acceptance and the fact that the target value for the fake-hit rate was \mbox{10$^{-6}$ hits/pixel/event} \cite{ITSUPTDR}. The result is consistent with what was observed with single chip characterisation.

Due to the very low recorded chip noise, hits coming from cosmic particles passing through the detector can be clustered together and used to reconstruct the associated tracks. Starting from tracks one can perform an alignment study to determine the displacement of staves in layers. Preliminary analysis gives residuals of the order of \mbox{100 $\mu$m}. The code developed using this alignment study can also be used to determine the final alignment, once the ITS 2 will be installed in the cavern, within the ALICE detector. The overlap of the staves at the edges facilitate the reconstruction of a track from 6 aligned clusters in three layers, as seen in the right plot of Fig. \ref{fig-4}.

Quality selection criterion to consider a stave good for installation in the IB was to have a total amount of not working pixels below 50 over the 9 chips. After installation, the total number of pixels found to be dead were less than 1000 out of a total \mbox{226 $\times$ 10$^{6}$} pixels of the full IB. This number is actually reduced with respect to the measurements done on the staves before installation in the barrel, thanks to improved stuck pixel identification and masking mechanism.   

Readout tests are used to verify the quality of the data transmission from the chips to the associated RU performing a measurement of the Bit Error Rate (BER) through counting of 8b10b decoding errors. During the test, a fixed number of pixels are stimulated and the chip matrix is read out at a given rate. Multiple injected patterns (256, 512, 1008, 2048 and 4096 pixels) and readout rates (\mbox{44.9 kHz}, \mbox{67.3 kHz}, \mbox{101 kHz}, \mbox{123 kHz}, \mbox{202 kHz} and \mbox{247 kHz}) were explored. From \cite{ITSUPTDR}, a total of $\sim$250 activated pixels are expected in case of \mbox{100 kHz} \mbox{Pb--Pb} minimum-bias collisions, adding QED background and pixel noise of the order of \mbox{10$^{-5}$ hits/pixel/event}, for an integration time of \mbox{30 $\mu$s}. Simulating an occupancy similar to the one described, a BER sensitivity of about 10$^{-12}$ was measured for all the readout rates mentioned above. The BER sensitivity was measured to be of the order of 10$^{-15}$ in a large statistics sample for the readout rate of 44.9 kHz. In a few extreme conditions, actually not corresponding to possible scenarios during data taking with beams, errors have been recorded on a limited amount of staves and with a rate of error of the order of $\sim$20 hours. Exploration of these extreme conditions as well as repetition of campaign with back-bias voltage at -3 V and investigation of boundary behaviour varying the digital supply voltage at the chip are in progress at the moment. 

\subsubsection{Outer Barrel}
At the arrival at CERN, before the final installation in the detector, each OB stave went through basic tests consisting of verification of the power consumption, the control communication and the high speed links.

Handling of very long and fragile objects, as the staves during the assembly of the layers and half-barrel, required the definition of a rigorous and yet delicate procedure. In spite of the complexity of the stave installation procedure, there was no increase in the number of dead chips, as ascertained from electrical and functional checks. The total number of dead chips over the full OB is 32, corresponding to the 0.14\% of the acceptance; no overlap in the radial direction is present between the dead chips in the different layers. 

\begin{figure}[t]
\centering
\includegraphics[width=37pc,clip]{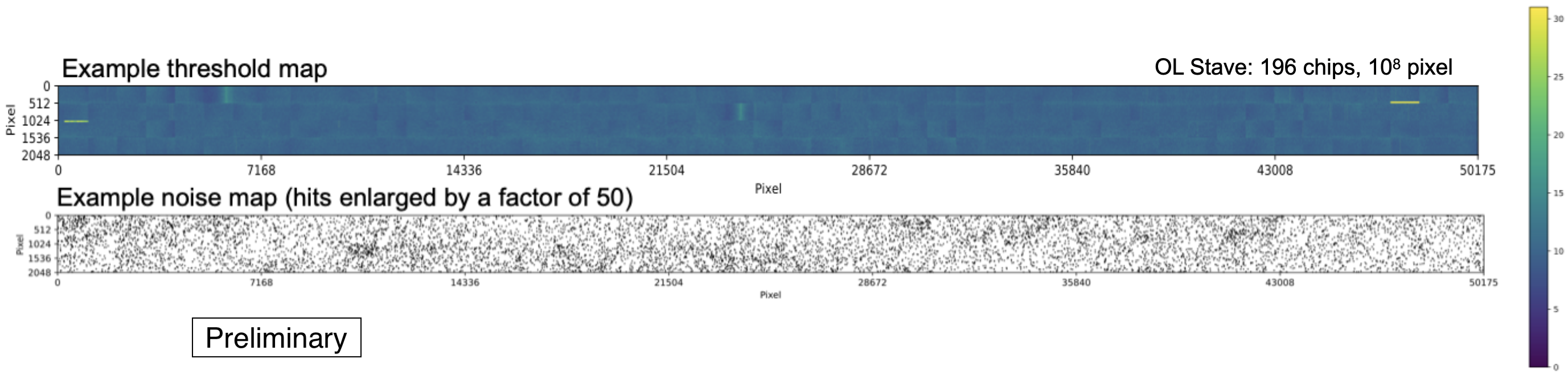}
\caption{Outer Barrel stave threshold map (top) and fake-hit rate map (bottom). Colour scale for the threshold value is reported on the right. Points in the fake-hit rate map are enlarged by a factor of 50 to make them visible.}
\label{fig-6}
\end{figure}

A calibration campaign allowed to characterise the OB staves, and showed that a tuning to \mbox{100 e$^{-}$} threshold target with a \mbox{2 e$^{-}$} precision is possible, having a chip-to-chip spread within the stave of \mbox{20 e$^{-}$}. An example of threshold map for a full OB stave, containing 196 chips, is shown in top plot of Fig. \ref{fig-6}. 

Due to the very large amount of chips installed in the OB staves ($\sim$24 $\times$ 10$^{3}$ chips), the selection criteria were released with respect to the chip used in the IB. As a consequence we do expect reduced performance in terms of noise. Nonetheless a preliminary study of the fake-hit rate as a function of the number of the noisier masked pixel, as done for the IB, demonstrated that a fake-hit rate value of the order of \mbox{10$^{-10}$ hits/pixel/event} is reachable masking up to the 0.014\% of the full OB acceptance. This value is well below the requirement of fake-hit rate to be less than \mbox{10$^{-6}$ hits/pixel/event} \cite{ITSUPTDR}. An example of noise map for an OB stave, with points enlarged by a factor 50 to make them visible, is shown in bottom plot of Fig. \ref{fig-6}. 

A long term continuous powering campaign, to bring out events of early infant dead of staves and any service components (e.g. power supply modules), was carried out and no suspicious cases have been found. This campaign required the improvement of the monitoring processes of a lot of detector components. 

Next steps in the OB staves commissioning, foresee a chip to RU communication verification campaign, as done for the IB, and collection of tracks from cosmic muons to allow alignment study. The first data collected using cosmic rays was used to improve the code for the track reconstruction.

\section{Conclusions}
The ongoing replacement of the ITS will extend the physics reach of ALICE to lower transverse momentum, allowing the characterisation of the QGP via measurements of unprecedented precision. 
The new ITS is now fully constructed after a huge effort. The working of readout chain for both IB and OB are demonstrated. IB is fully characterised and boundary limits are under exploration. The characterisation of the OB is ongoing and a campaign to gather particle tracks from cosmic muons is ready to begin.
The commissioning at the surface will continue until the end of 2020. In the beginning of 2021 the detector will be prepared to be transferred to ALICE P2. The installation in the cavern, testing of functionalities and other procedures will take about two months. After that, ITS 2 will join the other detectors in the global ALICE commissioning, in order to be ready to take the very first data with collisions in the foreseen LHC pilot run during October 2021.


\end{document}